\documentclass[prd,aps,floatfix,nofootinbib,11 pt]{revtex4}
\usepackage{amsmath,graphicx,color,epsfig}

\input{tcilatex}

\begin{document}

\title{Noisy non-transitive quantum games}
\author{M. Ramzan\thanks{%
mramzan@phys.qau.edu.pk}, Salman Khan and M. Khalid Khan}
\address{Department of Physics Quaid-i-Azam University \\
Islamabad 45320, Pakistan}

\date{\today }

\begin{abstract}
We study the effect of quantum noise in $3\times 3$ entangled quantum games.
By considering different noisy quantum channels we analyze that how a
two-player, three-strategy Rock-Scissor-Paper game is influenced by the
quantum noise. We consider the winning non-transitive strategies $R$, $S$
and $P$ such as $R$ beats $S$, $S$ beats $P$, and $P$ beats $R$. The game
behaves as a noiseless game for maximum value of the quantum noise
parameter. It is seen that Alice's payoff is heavily influenced by the
depolarizing noise as compared to the amplitude damping noise. Depolarizing
channel causes a monotonic decrease in players payoffs as we increase the
amount of of quantum noise. In case of amplitude damping channel, the
Alice's payoff function reaches its minimum for $\alpha =0.5$ and is
symmetrical. This means that larger values of quantum noise influence the
game weakly. On the other hand, phase damping channel does not influence the
game's payoff. Furthermore, the game's Nash equilibrium and non-transitive
character of the game are not affected under the influence of quantum noise.%
\newline
\end{abstract}

\maketitle
\date{\today}

\address{Department of Physics Quaid-i-Azam University \\
Islamabad 45320, Pakistan}


\vspace*{1.0cm}

\vspace*{1.0cm}


\section{Introduction}

In the recent past, rapid interest has been developed in the discipline of
quantum information \cite{NieMA} that has led to the creation of quantum
game theory [2-5]. During last few years, tremendous efforts have been made
towards the development of quantum game theory [6-12]. In this connection,
much work has been devoted to convert the classical games into quantum
domain such as prisoners' dilemma game [13-15] and many other games [16-20]
involving two and three players. Quantum games with $3\times 3$ payoff
matrices and larger have have been discussed by Wang et al. [21]. Recently,
Sousa et al. [22] have proposed that quantum games can be used to control
the access of processes to the CPU in a quantum computer. More recently,
Iqbal and Abbott [23] have reported that quantum games can directly be
constructed from a system of Bell's inequalities using Arthur Fine's
analysis.

In quantum information processing, the major problem is to faithfully
transmit unknown quantum states through a noisy quantum channel. When
quantum information is sent through a channel, the carriers of the
information interact with the channel and get entangled with its many
degrees of freedom. This gives rise to the phenomenon of decoherence on the
state space of the information carriers. In real-world applications, the
decoherence effects caused by the external environment are inevitable.
Quantum channels [1] provide a natural theoretical framework for the study
of decoherence in noisy quantum communication systems. Quantum games in the
presence of decoherence have been discussed in the recent past by various
authors [5, 24-26]. Recently, decoherence and correlated noise (memory)
effects have been analyzed in different quantum games [8, 20, 27]. Quantum
error correction [28-29] and entanglement purifications [30].can be employed
to avoid the problem of decoherence.

In this paper, we study the effect of quantum noise in a 2-player 3-strategy
entangled quantum game ($RSP$ game). We consider different noisy channels
parameterized by a quantum noise parameter $\alpha $ such that $\alpha \in
\lbrack 0,1]$. The lower and upper limits of quantum noise parameter
represent the fully coherent and fully decohered systems, respectively. It
is seen that for maximum value of quantum noise parameter the game behaves
as a noiseless game. The depolarizing channel influences the game's payoff
more heavily as compared to the amplitude damping channel. The payoff
function for amplitude damping channel, is symmetrical with its minimum at $%
\alpha =0.5$. Furthermore, the phase damping channel does not influence the
game.

\section{Noisy quantum $RSP$ game}

The rock, scissors and paper game is a game for two players typically played
using the players' hands. It is a simple two-player, three-strategy game,
the children's choosing game \textquotedblleft rock ($R$), scissors ($S$)
and paper ($P$)\textquotedblright\ denoted by $RSP$, in which rock beats
scissors, scissors beats paper, and paper beats rock $(R>S>P>R)$. The
classical payoff matrix for this game is given in table 1. Since $RSP$ game
is a zero-sum game, therefore it has no pure strategy Nash equilibrium.
However, the mixed strategy Nash equilibrium produces zero expected payoff
for the classical game.

Quantum analog of the $RSP$ game was formulated by Iqbal et al. [12] and
Michael et al. [31]. In the quantum version of this game, the strategies $R$%
, $S$, and $P$ are represented by three matrices ($U_{1}$, $U_{2}$ and $%
U_{3} $ equation 9 Ref. [31]). $U(x,y),$ is the unitary matrix which is used
to transform the initial qutrit state of the game. We assume that initially
Alice and Bob share a maximally entangled two qutrit state of the form

\begin{equation}
|\Psi _{in}\rangle =\frac{1}{\sqrt{3}}\left( |00\rangle +|11\rangle
+|22\rangle \right)  \label{init}
\end{equation}%
We can define the strategies of the players by the unitary operator $U(x,y)$
of the form [31] 
\begin{equation}
U(x,y)=\frac{1}{\sqrt{2}}\left[ 
\begin{array}{ccc}
e^{ix}\cos y & ie^{ix}\sin y & 0 \\ 
i\sin y\cos x & \cos x\cos y & ie^{iy}\sin x \\ 
-\sin y\sin x & i\sin x\cos y & e^{iy}\cos x%
\end{array}%
\right]
\end{equation}%
where $0\leq \{x,$ $y\}\leq \pi /2$ and the choice of $x$ and $y$ define a
player's mixed strategy in a $3\times 3$ game, just as the choice of $\theta 
$ does in the $2\times 2$ game [3].

The interaction between the system and its environment introduces the
decoherence to the system, which is a process of the undesired correlation
between the system and the environment when the system evolves. The
evolution of a state of a quantum system in a noisy environment can be
described by the super-operator $\Phi $ in the Kraus operator representation
[1] as

\begin{equation}
\rho _{f}=\Phi \rho _{i}=\sum_{k}E_{k}\rho _{i}E_{k}^{\dag }  \label{E5}
\end{equation}%
where the Kraus operators $E_{i}$ satisfy the following completeness relation

\begin{equation}
\sum_{k}E_{k}^{\dag }E_{k}=I  \label{5}
\end{equation}%
We have constructed the Kraus operators for the game from the single qutrit
Kraus operators (as given in equations (6-9) below) by taking their tensor
product over all $n^{2}$ combination of $\pi \left( i\right) $ indices

\begin{equation}
E_{k}=\underset{\pi }{\otimes }e_{\pi \left( i\right) }  \label{6}
\end{equation}%
where $n$ is the number of Kraus operators for a single qutrit channel. The
single qutrit Kraus operators for the amplitude damping channel are given by
[32]

\begin{equation}
E_{0}=\left( 
\begin{array}{ccc}
1 & 0 & 0 \\ 
0 & \sqrt{1-\alpha } & 0 \\ 
0 & 0 & \sqrt{1-\alpha }%
\end{array}%
\right) ,\ \ E_{1}=\left( 
\begin{array}{ccc}
0 & \sqrt{\alpha } & 0 \\ 
0 & 0 & 0 \\ 
0 & 0 & 0%
\end{array}%
\right) ,\ \ E_{2}=\left( 
\begin{array}{ccc}
0 & 0 & \sqrt{\alpha } \\ 
0 & 0 & 0 \\ 
0 & 0 & 0%
\end{array}%
\right)  \label{E7}
\end{equation}%
Similarly, the single qutrit Kraus operators for the phase damping channel
are given as [32]

\begin{equation}
E_{0}=\sqrt{1-\alpha }\left( 
\begin{array}{ccc}
1 & 0 & 0 \\ 
0 & 1 & 0 \\ 
0 & 0 & 1%
\end{array}%
\right) ,\ \ E_{1}=\sqrt{\alpha }\left( 
\begin{array}{ccc}
1 & 0 & 0 \\ 
0 & \omega & 0 \\ 
0 & 0 & \omega ^{2}%
\end{array}%
\right) ,  \label{7}
\end{equation}%
The single qutrit Kraus operators for the depolarizing channel are given by
[33]

\begin{equation*}
E_{0}=\sqrt{1-\alpha }I,\ E_{1}=\sqrt{\frac{\alpha }{8}}Y,\ E_{2}=\sqrt{%
\frac{\alpha }{8}}Z,\ E_{3}=\sqrt{\frac{\alpha }{8}}Y^{2},\ E_{4}=\sqrt{%
\frac{\alpha }{8}}YZ
\end{equation*}

\begin{equation}
E_{5}=\sqrt{\frac{\alpha }{8}}Y^{2}Z,\ E_{6}=\sqrt{\frac{\alpha }{8}}%
YZ^{2},\ \ E_{7}=\sqrt{\frac{\alpha }{8}}Y^{2}Z^{2},\ \ E_{8}=\sqrt{\frac{%
\alpha }{8}}Z^{2}  \label{E8}
\end{equation}%
where

\begin{equation}
Y=\left( 
\begin{array}{ccc}
0 & 1 & 0 \\ 
0 & 0 & 1 \\ 
1 & 0 & 0%
\end{array}%
\right) ,\ \ Z=\left( 
\begin{array}{ccc}
1 & 0 & 0 \\ 
0 & \omega & 0 \\ 
0 & 0 & \omega ^{2}%
\end{array}%
\right)  \label{9}
\end{equation}%
In the above equations $\alpha $ represents the quantum noise parameter and $%
\omega =e^{\frac{2\pi i}{3}}$. The final state of the game after the action
of the channel can be written as 
\begin{equation}
\rho _{f}=\Phi _{\alpha }(\left\vert \Psi _{in}\right\rangle \left\langle
\Psi _{in}\right\vert )
\end{equation}%
where $\Phi _{\alpha }$ is the super-operator realizing the quantum channel
parametrized by the real number $\alpha $ (quantum noise parameter). After
the action of the players unitary operations, the game's final state
transforms to 
\begin{equation}
\rho _{\acute{f}}=(U_{A}(x,y)\otimes U_{B}(x,y))(\Phi _{\alpha }(\left\vert
\Psi _{in}\right\rangle \left\langle \Psi _{in}\right\vert )(U_{A}^{\dag
}(x,y)\otimes U_{B}^{\dag }(x,y))  \label{rf}
\end{equation}%
The payoff operators for Alice and Bob can be written as 
\begin{eqnarray}
(P^{A,B})_{\text{Oper.}}
&=&\$_{00}^{A,B}P_{00}+\$_{01}^{A,B}P_{01}+\$_{02}^{A,B}P_{02}+%
\$_{10}^{A,B}P_{10}+\$_{11}^{A,B}P_{11}  \notag \\
&&+\$_{12}^{A,B}P_{12}+\$_{20}^{A,B}P_{20}+\$_{21}^{A,B}P_{21}+%
\$_{22}^{A,B}P_{22}
\end{eqnarray}%
where $\$_{ij}^{A,B}$ are the elements of payoff matrix in the $i$th row and 
$j$th column of classical game as given in table 1 and 
\begin{eqnarray}
P_{00} &=&\left\vert 00\right\rangle \left\langle 00\right\vert ,\text{ }%
P_{01}=\left\vert 01\right\rangle \left\langle 01\right\vert ,\text{ }%
P_{02}=\left\vert 02\right\rangle \left\langle 02\right\vert  \notag \\
P_{10} &=&\left\vert 10\right\rangle \left\langle 10\right\vert ,\text{ }%
P_{11}=\left\vert 11\right\rangle \left\langle 11\right\vert ,\text{ }%
P_{12}=\left\vert 12\right\rangle \left\langle 12\right\vert  \notag \\
P_{20} &=&\left\vert 20\right\rangle \left\langle 20\right\vert ,\text{ }%
P_{21}=\left\vert 21\right\rangle \left\langle 21\right\vert ,\text{ }%
P_{22}=\left\vert 22\right\rangle \left\langle 22\right\vert
\end{eqnarray}%
The players payoffs can be calculated by using the relation 
\begin{equation}
\$^{A,B}(x_{i},y_{i},\alpha )=\text{Tr\{[}(P^{A,B})_{\text{Oper.}}\rho
_{f}]^{A,B}\}
\end{equation}%
where Tr represents the trace of the matrix.

\subsection{Results using different noise models}

By using equations (1-6, 11-13 and 17), the players payoffs for the
amplitude damping channel can be obtained as%
\begin{eqnarray}
\$_{\text{AD}}^{A,B}(x_{i},y_{i},\alpha ) &=&\frac{1}{12}[(2+\alpha
^{2})\$_{00}+\{2+\alpha (-4+5\alpha )\}\$_{00}\cos (2y_{1})\cos (2y_{2}) 
\notag \\
&&+\alpha (2+\alpha )\$_{00}(\cos (2y_{1})+\cos (2y_{2}))+2(-1+\alpha )\times
\notag \\
&&[(-\$_{11}+\$_{12}+\$_{21}-\$_{22})\cos (y_{1}+y_{2})  \notag \\
&&\sin (2x_{1})\sin (2x_{2})\{(-1+\alpha )\cos (2y_{1})\cos (2y_{2})+\sin
(y_{1})\sin (y_{2})\}  \notag \\
&&+\$_{00}\sin (2y_{1})\sin (2y_{2})]+\cos (x_{1})^{2}[-4(-1+\alpha )\alpha
\$_{20}\cos (y_{2})^{2}  \notag \\
&&-\$_{10}\cos (2y_{1})\{\alpha (2+\alpha )+(2+\alpha (-4+5\alpha ))\cos
(2y_{2})\}  \notag \\
&&+\$_{10}\{2+\alpha ^{2}+\alpha (2+\alpha )\cos (2y_{2})-2(-1+\alpha )\sin
(2y_{1})\sin (2y_{2})\}]  \notag \\
&&+\cos (x_{2})^{2}[(2+\alpha ^{2})\$_{01}+\alpha (2+\alpha )\$_{01}(\cos
(2y1)-\cos (2y_{2}))  \notag \\
&&+\{-2+(4-5\alpha )\alpha \}\$_{01}\cos (2y_{1})\cos (2y_{2})-2(-1+\alpha
)(2\alpha \$_{02}\cos (y_{1})^{2}  \notag \\
&&+\$_{01}\sin (2y_{1})\sin (2y_{2}))+\cos (x_{1})^{2}\{(2+\alpha
^{2})\$_{11}+4(-1+\alpha )^{2}\$_{22}  \notag \\
&&+(2+\alpha (-4+5\alpha ))\$_{11}\cos (2y_{1})\cos (2y_{2})-\alpha
(2+\alpha )\$_{11}(\cos (2y_{1})+\cos (2y_{2}))  \notag \\
&&-2(-1+\alpha )(2\alpha \$_{12}\sin (y_{1})^{2}+2\alpha \$_{21}\sin
(y_{2})^{2}-\$_{11}\sin (2y_{1})\sin (2y_{2}))\}]  \notag \\
&&+\sin (x_{2})^{2}[-4(-1+\alpha )\alpha \$_{01}\cos
(y_{1})^{2}+\$_{02}\{2+\alpha ^{2}-\alpha (2+\alpha )\cos (2y_{2})  \notag \\
&&+\cos (2y_{1})[\alpha (2+\alpha )+\{-2+(4-5\alpha )\alpha \}\cos
(2y_{2})]-2(-1+\alpha )\sin (2y_{1})\sin (2y_{2})\}  \notag \\
&&+\cos (x_{1})^{2}\{(2+\alpha ^{2})\$_{12}+4(-1+\alpha )^{2}\$_{21}-\alpha
(2+\alpha )\$12\cos (2y_{2})  \notag \\
&&+\$_{12}\cos (2y_{1})[-\alpha (2+\alpha )+\{2+\alpha (-4+5\alpha )\}\cos
(2y_{2})]  \notag \\
&&-2(-1+\alpha )(2\alpha \$_{11}\sin (y_{1})^{2}+2\alpha \$_{22}\sin
(y_{2})^{2}-\$_{12}\sin (2y_{1})\sin (2y_{2}))\}]  \notag \\
&&+\sin (x_{1})^{2}[2\$_{20}+\alpha ^{2}\$_{20}+4\alpha \$_{10}\cos
(y_{2})^{2}-4\alpha ^{2}\$_{10}\cos (y_{2})^{2}  \notag \\
&&-\alpha (2+\alpha )\$_{20}(\cos (2y_{1})-\cos (2y_{2}))+\{-2+(4-5\alpha
)\alpha \}\$_{20}\cos (2y_{1})\cos (2y_{2})  \notag \\
&&-2(-1+\alpha )\$_{20}\sin (2y_{1})\sin (2y_{2})+2\cos
(x_{2})^{2}[2(-1+\alpha )^{2}\$_{12}  \notag \\
&&+2\sin (y_{1})^{2}\{-(-1+\alpha )\alpha (\$_{22}+\$_{21}\cos
(y_{2})^{2})+(1+2\alpha ^{2})\$_{21}\sin (y_{2})^{2}\}  \notag \\
&&+(-1+\alpha )[\$_{21}\cos (y_{1})^{2}\{-1+(-1+2\alpha )\cos
(2y_{2})\}-2\alpha \$_{11}\sin (y_{2})^{2}  \notag \\
&&+\$_{21}\sin (2y_{1})\sin (2y_{2})]]+2\sin (x_{2})^{2}[(-1+\alpha
)\$_{22}\cos (y_{1})^{2}  \notag \\
&&\times \{-1+(-1+2\alpha )\cos (2y_{2})\}+\{\$_{22}+\alpha
(2\$_{21}-2\alpha \$_{21}+\$_{22}+\alpha \$_{22})  \notag \\
&&+(-1+\alpha -3\alpha ^{2})\$_{22}\cos (2y_{2})\}\sin
(y_{1})^{2}+(-1+\alpha )\{2(-1+\alpha )\$_{11}  \notag \\
&&-2\alpha \$_{12}\sin (y_{2})^{2}+\$_{22}\sin (2y_{1})\sin (2y_{2})\}]]]
\end{eqnarray}%
By using equations (1-5, 8-13 and 17), the players payoffs for the
depolarizing channel become%
\begin{eqnarray}
\$_{\text{Dep}}^{A,B}(x_{i},y_{i},\alpha ) &=&\frac{1}{3072}[(8-9\alpha
)^{2}(\$_{11}-\$_{12}-\$_{21}+\$_{22})\cos (2(x_{1}+x_{2}))  \notag \\
&&\times (5+3\cos (2(y_{1}+y_{2})))+2\{4\{64+3\alpha (-16+9\alpha )\}\$_{00}
\notag \\
&&-3\alpha (-16+9\alpha )(2\$_{01}+2\$_{02}+2\$_{10}-\$_{11}-\$_{12}+2\$_{20}
\notag \\
&&-\$_{21}-\$_{22})+64\{2\$_{01}+2\$_{02}+2\$_{10}+3\$_{11}+3\$_{12}  \notag
\\
&&+2\$_{20}+3(\$_{21}+\$_{22})\}+(8-9\alpha )^{2}\{(4\$_{00}-2\$_{01}  \notag
\\
&&-2\$_{02}-2\$_{10}+\$_{11}+\$_{12}-2\$_{20}+\$_{21}+\$_{22})\cos
(2(y_{1}+y_{2}))  \notag \\
&&+\{2(2\$_{10}-\$_{11}-\$_{12}-2\$_{20}+\$_{21}+\$_{22})\cos (2x_{1}) 
\notag \\
&&+(\$_{11}-\$_{12}-\$_{21}+\$_{22})\cos (2(x_{1}-x_{2}))+  \notag \\
&&2(2\$_{01}-2\$_{02}-\$_{11}+\$_{12}-\$_{21}+\$_{22})\cos (2x_{2})\}\sin
(y_{1}+y_{2})^{2}\}\}]  \notag \\
&&
\end{eqnarray}%
By using equations (1-5, 7, 10-13 and 17), the players payoffs for the phase
damping channel are obtained as%
\begin{eqnarray}
\$_{\text{PD}}^{A,B}(x_{i},y_{i},\alpha ) &=&\frac{1}{192}%
[-8(2\$_{11}+2\$_{12}+2\$_{20}+\$_{21}+\$_{22})\cos (2x_{1})  \notag \\
&&+\cos (2x_{2})[8\{2\$_{01}-2\$_{02}-2\$_{11}+2\$_{12}+\$_{21}-\$_{22} 
\notag \\
&&+(2\$_{11}-2\$_{12}-\$_{21}+\$_{22})\cos (2x_{1})\}-8\{2\$_{01}-2\$_{02} 
\notag \\
&&-\$_{21}+\$_{22}+(\$_{21}-\$_{22})\cos (2x_{1})\}\cos (2y_{1})\cos (2y_{2})
\notag \\
&&+4\{2+3(-2+\alpha )\alpha \}\{2\$_{01}-2\$_{02}-\$_{21}+\$_{22}  \notag \\
&&+(\$_{21}-\$_{22})\cos (2x_{1})\}\sin (2y_{1})\sin (2y_{2})]  \notag \\
&&+16\cos (2x_{1})^{2}[2\$_{10}+\$_{11}+\$_{12}+2\$_{21}+2\$_{22}  \notag \\
&&+(\$_{11}-\$_{12}-2\$_{21}+2\$_{22})\cos (2x_{2})+\{-2\$_{10}  \notag \\
&&+\$_{11}+\$_{12}+(\$_{11}-\$_{12})\cos (2x_{2})\}\cos (2y_{1})\cos (2y_{2})
\notag \\
&&+\frac{1}{2}\{2+3(-2+\alpha )\alpha \}\{2\$_{10}-\$_{11}-\$_{12}  \notag \\
&&+(-\$_{11}+\$_{12})\cos (2x_{2})\}\sin (2y_{1})\sin (2y_{2})]  \notag \\
&&-8(\$_{11}-\$_{12}-\$_{21}+\$_{22})\sin (2x_{1})\sin
(2x_{2})[2+3(-2+\alpha )\alpha  \notag \\
&&+\{2+3(-2+\alpha )\alpha \}\cos (2(y_{1}+y_{2})])+\sqrt{3}\alpha
(-2+3\alpha )(\sin (2y_{1})  \notag \\
&&+\sin
(2y_{2}))]+8[4\$_{00}+2\$_{01}+2\$_{02}+2\$_{11}+2\$_{12}+2\$_{20}+\$_{21} 
\notag \\
&&+\$_{22}+\{2\$_{00}-\$_{01}-\$_{02}+(-2\$_{20}+\$_{21}+\$_{22})\sin
(2x_{1})^{2}\}  \notag \\
&&\times \{2\cos (2y_{1})\cos (2y_{2})+\{-2-3(-2+\alpha )\alpha \}\sin
(2y_{1})\sin (2y_{2})\}]]  \notag \\
&&
\end{eqnarray}%
where the subscripts AD, Dep and PD in equations (18-20) represent the
amplitude damping, depolarizing and phase damping channels respectively. It
can be easily checked from equations (18-20) that by setting $\alpha =1,$
the game becomes a noiseless game.

\section{Discussions}

In this work, we analyze the non-transitive two-player three-strategy
entangled quantum game usually termed as $RSP$ game under the influence of
quantum noise. We consider different noisy quantum channels and show that
that how the game's payoff is influenced by these channels. We consider the
winning non-transitive strategies $R$, $S$ and $P$ such as $R$ beats $S$, $S$
beats $P$, and $P$ beats $R$ $(R>S>P>R)$. It is seen that the game's payoff
is differently influenced by different quantum channels.

In figure 1, we plot the Alice's payoff as a function of quantum noise
parameter $\alpha $ for $x_{1}=y_{1}=\pi /2,$ $x_{2}=y_{2}=0$ for amplitude
damping (solid line), depolarizing (dashed line) and phase damping (dotted
line) channels. It is seen that Alice's payoff is heavily influenced by
depolarizing noise as compared to the amplitude damping noise. It causes a
monotonic decrease in players payoffs as the amount of quantum noise is
increased. It is evident from the figure that in case of amplitude damping
channel, the Alice's payoff reaches its minimum for $\alpha =0.5$ and is
symmetrical. This implies that the larger amount of quantum noise influences
the game weakly. However, the phase damping channel has no influence on the
payoffs of the players.

In figure 2, we plot the Alice's payoff as a function of her strategy $x_{1}$
for $y_{1}=\pi /2,$ $x_{2}=y_{2}=0$ and $\alpha =0.5$ for amplitude damping
(solid line), depolarizing (dashed line) and phase damping (dotted line)
channels. It can be seen from the figure that the phase damping channel does
not influence the Alice's payoff. On the other hand, the Alice's payoff is
strongly affected for amplitude damping and depolarizing channels.

In figure 3, we plot the Alice's payoff as a function of her strategy $y_{1}$
for $x_{1}=\pi /2,$ $x_{2}=y_{2}=0$ and $\alpha =0.5$ for amplitude damping
(solid line), depolarizing (dashed line) and phase damping (dotted line)
channels. It can be seen that the Alice's payoff is decreased due to the
presence of quantum noise for both the amplitude damping and depolarizing
channels. Whereas in case of phase damping channel, Alice's payoff remains
unaffected in the presence of quantum noise. In addition, it can be easily
checked that for maximum value of quantum noise parameter (i.e. at $\alpha
=1 $) the game behaves as a noiseless game. For $\alpha =1,$\ we obtain a
single curve for all the three channels (the phase damping channel curve in
figures 2 and 3). Hence, we can say that the game becomes a noiseless game
for maximum value of quantum noise.

In figures 4-6, we present the 3D graphs of Alice's payoff as a function of $%
\alpha $\ and her strategy $x_{1}$ for $y_{1}=\pi /2,$ $x_{2}=y_{2}=0$\ for
amplitude damping, depolarizing and phase damping channels respectively. One
can easily see that the depolarizing channel influence the game more
strongly as compared to the amplitude damping channel. From figures 4-6, it
can also be seen that the Nash equilibrium of the game does not change under
the influence of quantum noise. Furthermore, the non-transitive character of
the classical game is not affected by the quantum noise.

\section{Conclusions}

We study the influence of quantum noise on the Rock-Scissor-Paper ($RSP$)
game under different noise models. We consider the winning non-transitive
strategies $R$, $S$ and $P$ such as $R$ beats $S$, $S$ beats $P$, and $P$
beats $R$. Our investigations show that for maximum value of quantum noise
parameter the game behaves as a noiseless game. It is seen that game's
payoff is strongly influenced by the depolarizing noise as compared to the
amplitude damping noise. It is shown that under the influence of
depolarizing channel the players payoffs decrease monotonically as a
function of quantum noise. However, in case of amplitude damping channel,
the payoff function reaches its minimum for $\alpha =0.5$ and is
symmetrical. Therefore, amplitude damping channel influences the game weakly
for higher values of quantum noise. Furthermore, the phase damping channel
does not influence the game's payoff. Therefore the game deserves a careful
study during its implementation. It is also seen that the non-transitive
character of the classical game and the game's Nash equilibrium are not
affected by the quantum noise.\newline

{\huge Figures captions}\newline
\textbf{Figure 1}. Alice's payoff plotted as a function of quantum noise
parameter $\alpha $ for $x_{1}=y_{1}=\pi /2,$ $x_{2}=y_{2}=0$ for amplitude
damping (solid line), depolarizing (dashed line) and phase damping (dotted
line) channels.\newline
\textbf{Figure 2}. Alice's payoff plotted as a function of her strategy $%
x_{1}$ for $y_{1}=\pi /2,$ $x_{2}=y_{2}=0$ and $\alpha =0.5$ for amplitude
damping (solid line), depolarizing (dashed line) and phase damping (dotted
line) channels.\newline
\textbf{Figure 3}. Alice's payoff plotted as a function of her strategy $%
y_{1}$ for $x_{1}=\pi /2,$ $x_{2}=y_{2}=0$ and $\alpha =0.5$ for amplitude
damping (solid line), depolarizing (dashed line) and phase damping (dotted
line) channels.\newline
\textbf{Figure 4}. Alice's payoff plotted as a function of her strategy $%
x_{1}$ and $\alpha $ for $y_{1}=\pi /2,$ $x_{2}=y_{2}=0$ for amplitude
damping channel.\newline
\textbf{Figure 5}. Alice's payoff plotted as a function of her strategy $%
x_{1}$ and $\alpha $ for $y_{1}=\pi /2,$ $x_{2}=y_{2}=0$ for depolarizing
channel.\newline
\textbf{Figure 6}. Alice's payoff plotted as a function of her strategy $%
x_{1}$ and $\alpha $ for $y_{1}=\pi /2,$ $x_{2}=y_{2}=0$ for phase damping
channel.\newline
{\Huge Table Caption}\newline
\textbf{Table 1}. In the \textquotedblleft rock, scissors,
paper\textquotedblright\ game, a player can win regardless of the strategy
chosen by an opponent. The first number in each entry corresponds to Alice's
payoff and the second number corresponds to Bob. Winning strategies are
non-transitive in that $R>S>P>R.$ A payoff of $+1$ has been assigned to
winning, $-1$ to losing and $0$ for both in case of a tie.\newpage

\begin{figure}[tbp]
\begin{center}
\vspace{-2cm} \includegraphics[scale=0.6]{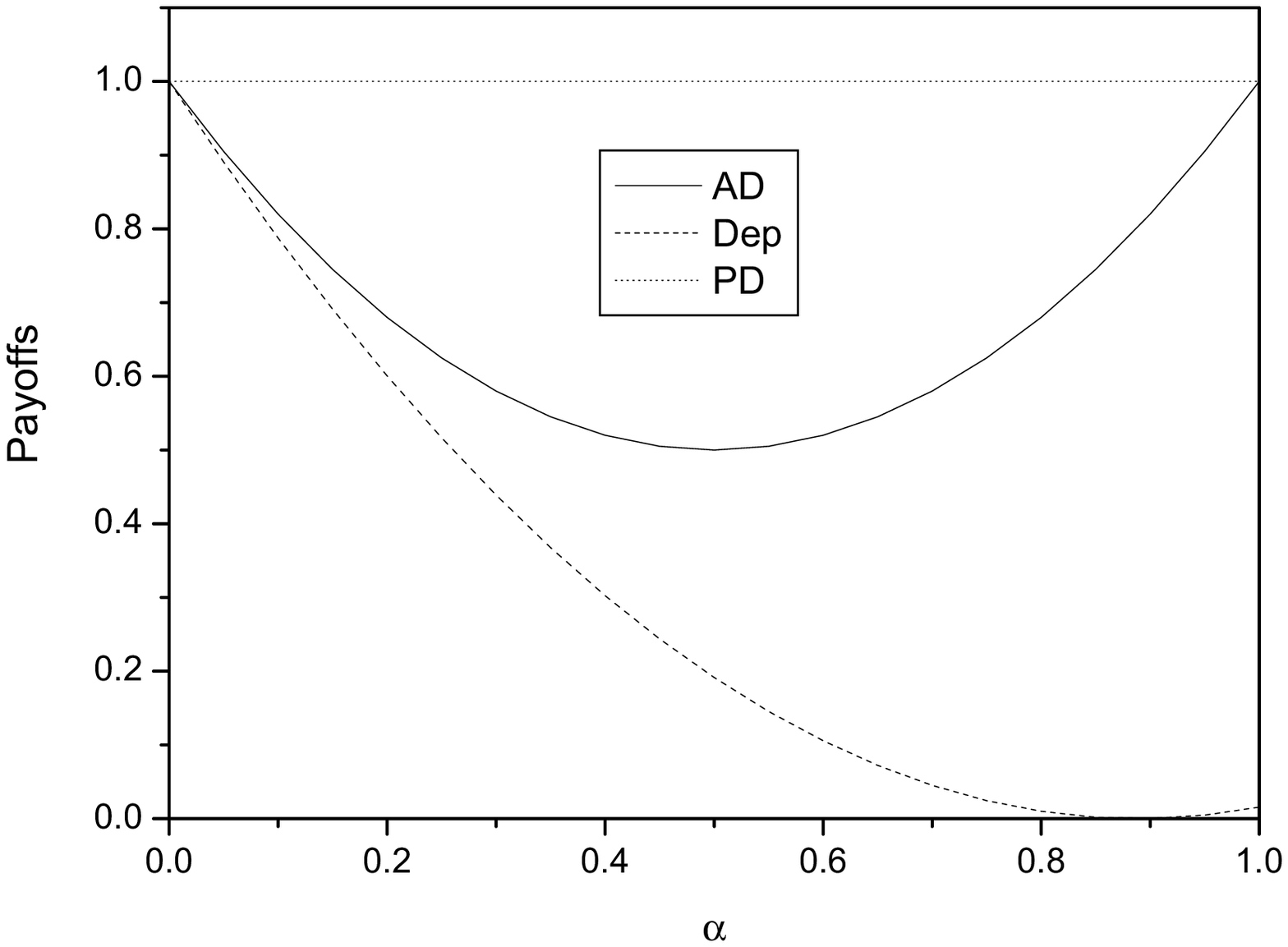} \\[0pt]
\end{center}
\caption{Alice's payoff plotted as a function of quantum noise parameter $%
\protect\alpha $ for $x_{1}=y_{1}=\protect\pi /2,$ $x_{2}=y_{2}=0$ for
amplitude damping (solid line), depolarizing (dashed line) and phase damping
(dotted line) channels.}
\end{figure}

\begin{figure}[tbp]
\begin{center}
\vspace{-2cm} \includegraphics[scale=0.6]{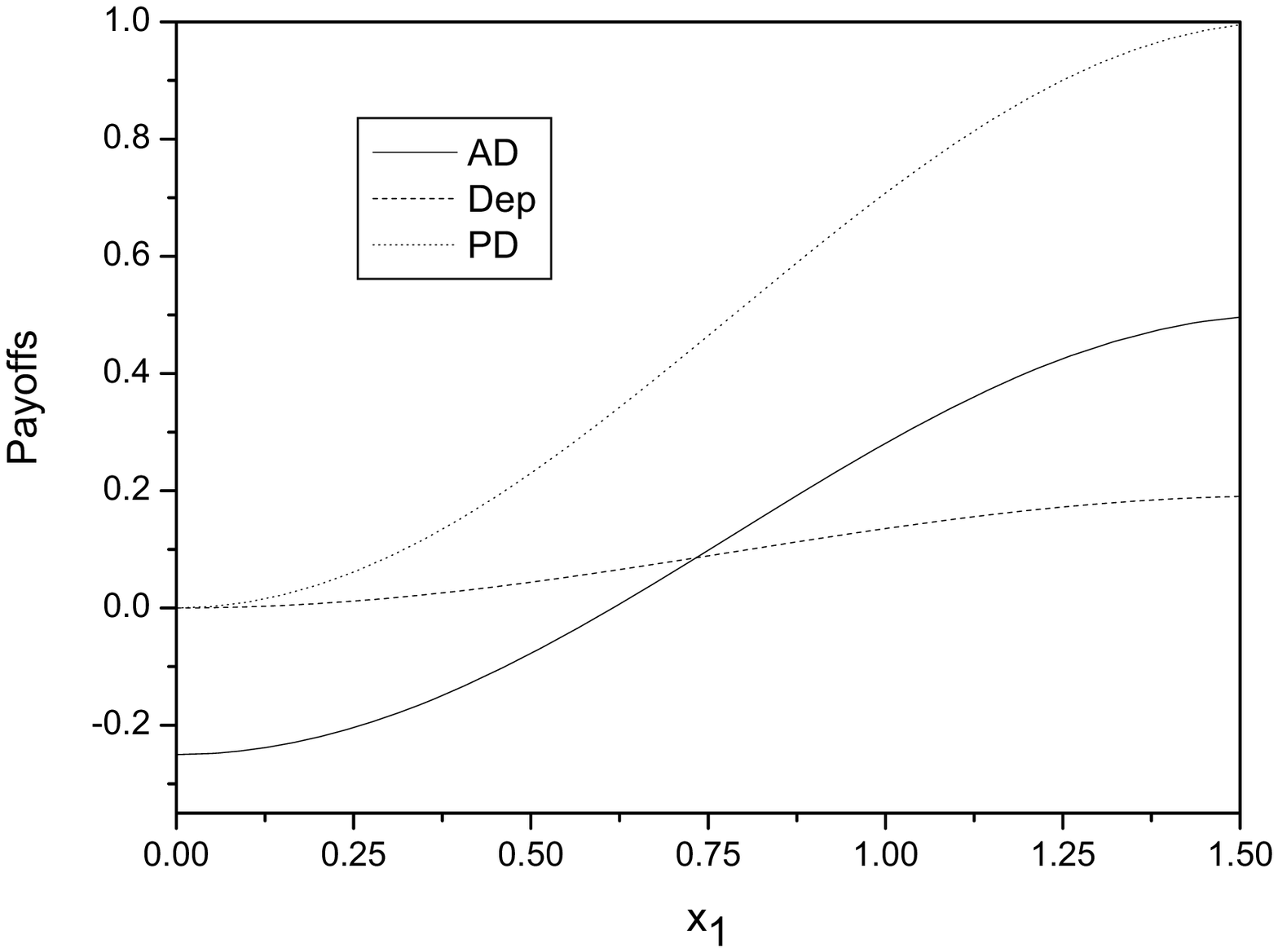} \\[0pt]
\end{center}
\caption{Alice's payoff plotted as a function of her strategy $x_{1}$ for $%
y_{1}=\protect\pi /2,$ $x_{2}=y_{2}=0$ and $\protect\alpha =0.5$ for
amplitude damping (solid line), depolarizing (dashed line) and phase damping
(dotted line) channels.}
\end{figure}

\begin{figure}[tbp]
\begin{center}
\vspace{-2cm} \includegraphics[scale=0.6]{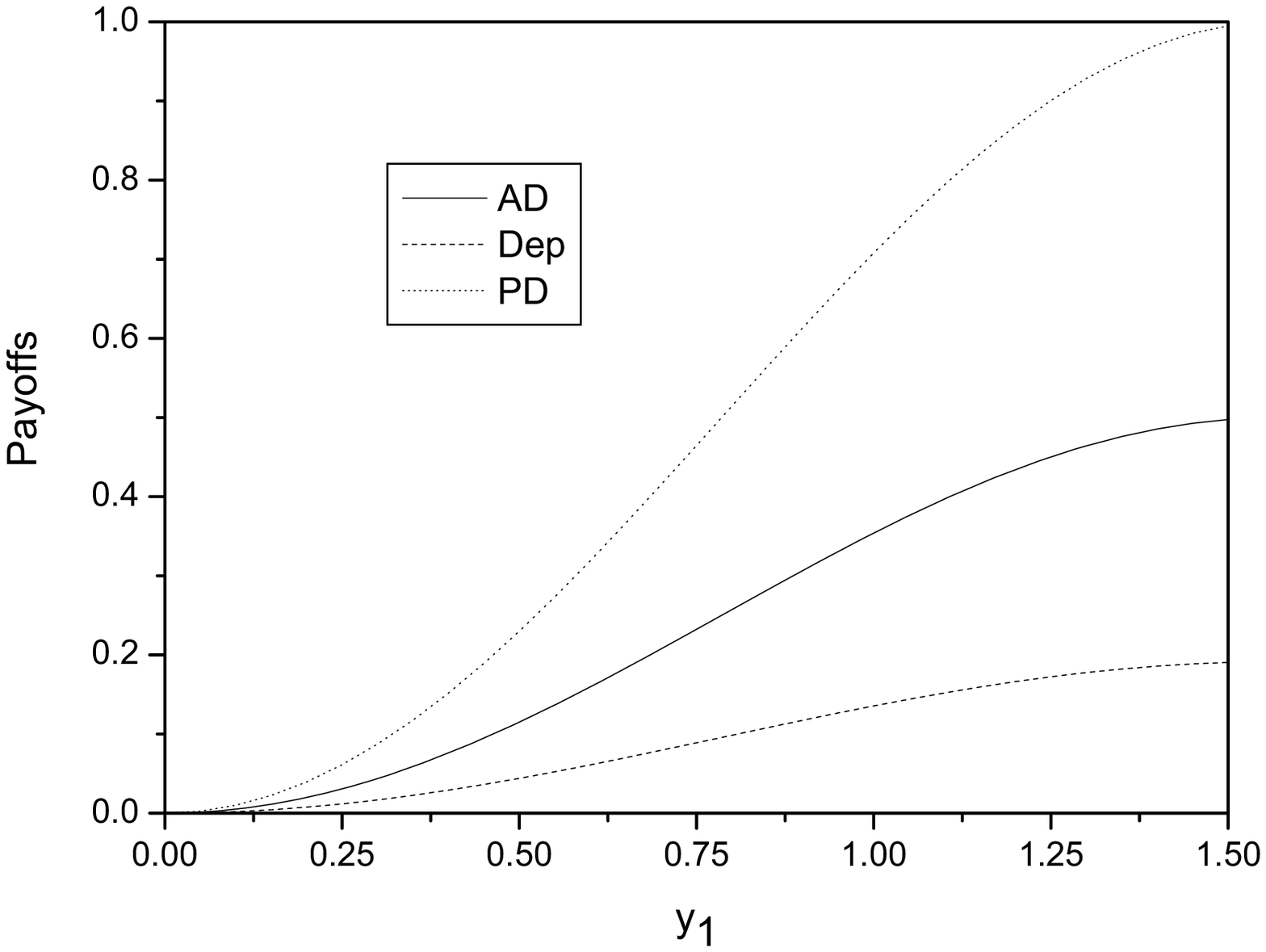} \\[0pt]
\end{center}
\caption{Alice's payoff plotted as a function of her strategy $y_{1}$ for $%
x_{1}=\protect\pi /2,$ $x_{2}=y_{2}=0$ and $\protect\alpha =0.5$ for
amplitude damping (solid line), depolarizing (dashed line) and phase damping
(dotted line) channels.}
\end{figure}

\begin{figure}[tbp]
\begin{center}
\vspace{-2cm} \includegraphics[scale=0.6]{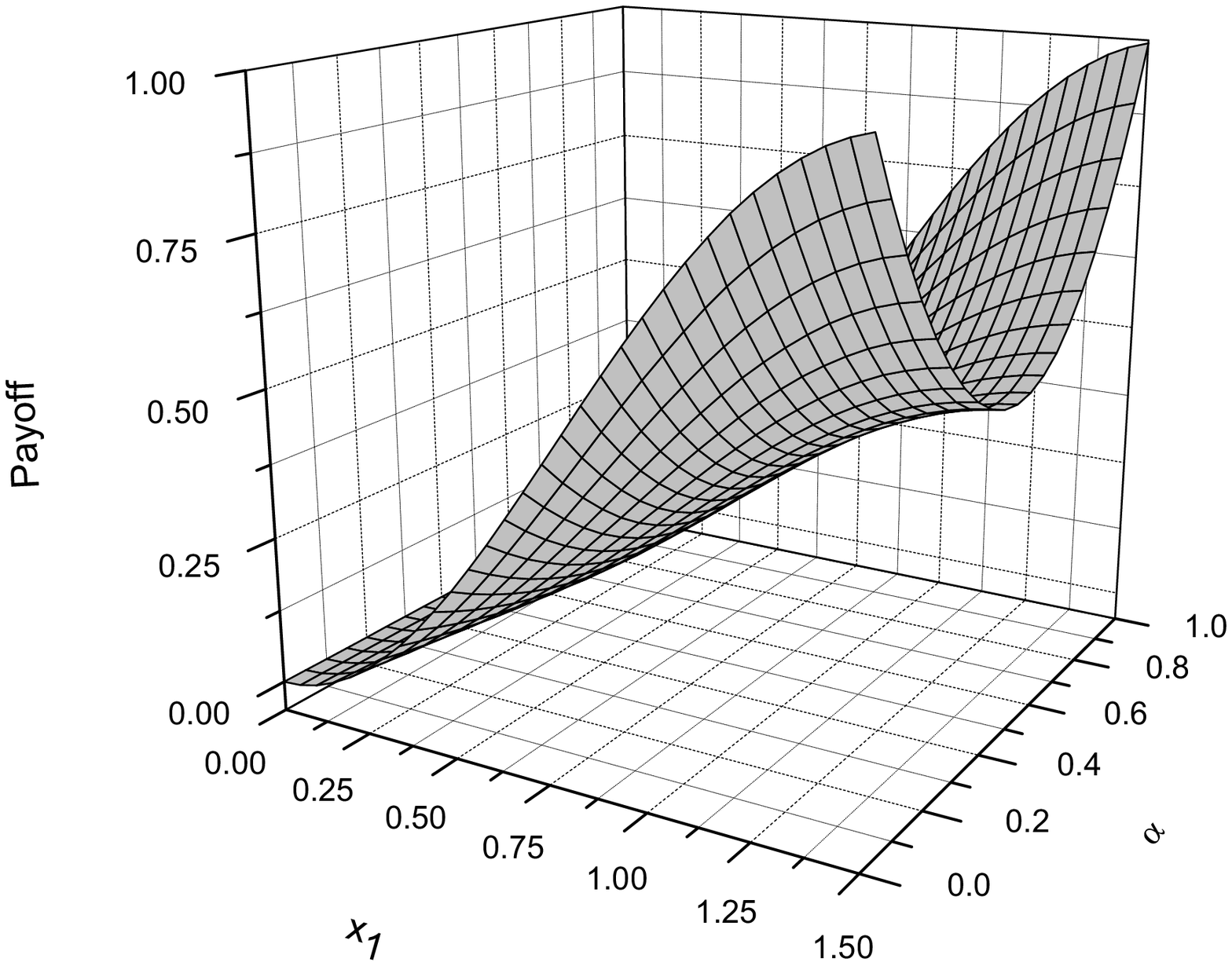} \\[0pt]
\end{center}
\caption{Alice's payoff plotted as a function of her strategy $x_{1}$ and $%
\protect\alpha $ for $y_{1}=\protect\pi /2,$ $x_{2}=y_{2}=0$ for amplitude
damping channel.}
\end{figure}

\begin{figure}[tbp]
\begin{center}
\vspace{-2cm} \includegraphics[scale=0.6]{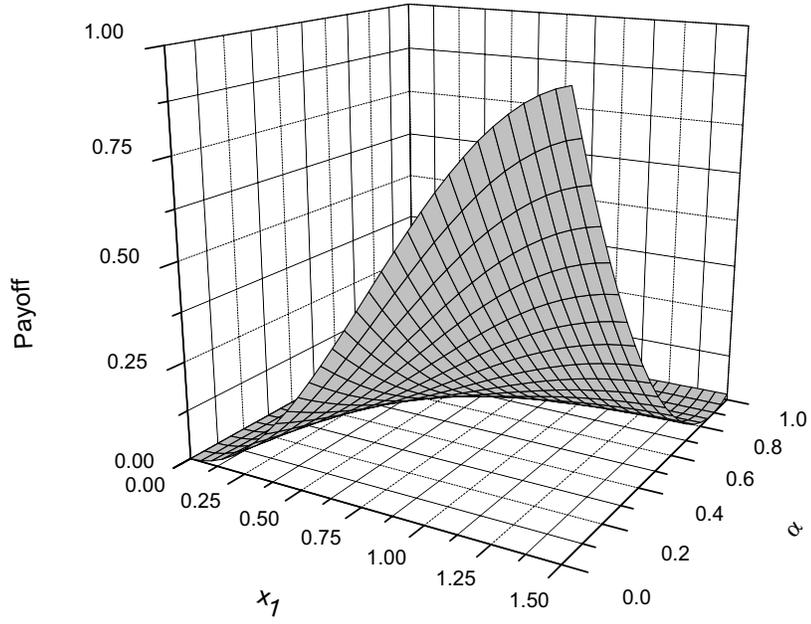} \\[0pt]
\end{center}
\caption{Alice's payoff plotted as a function of her strategy $x_{1}$ and $%
\protect\alpha $ for $y_{1}=\protect\pi /2,$ $x_{2}=y_{2}=0$ for
depolarizing channel.}
\end{figure}

\begin{figure}[tbp]
\begin{center}
\vspace{-2cm} \includegraphics[scale=0.6]{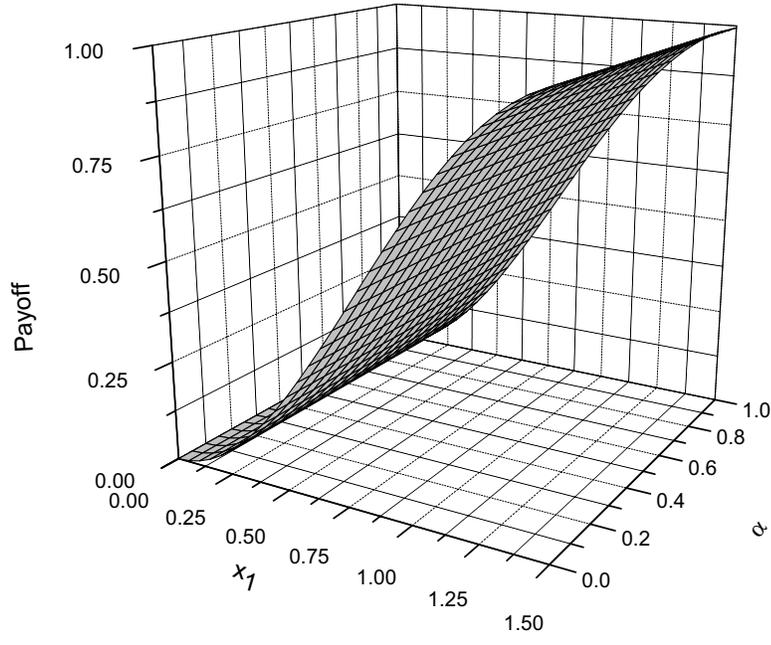} \\[0pt]
\end{center}
\caption{Alice's payoff plotted as a function of her strategy $x_{1}$ and $%
\protect\alpha $ for $y_{1}=\protect\pi /2,$ $x_{2}=y_{2}=0$ for phase
damping channel.}
\end{figure}
\begin{table}[tbh]
\caption{In the \textquotedblleft rock, scissors, paper\textquotedblright\
game, a player can win regardless of the strategy chosen by an opponent. The
first number in each entry corresponds to Alice's payoff and the second
number corresponds to Bob. Winning strategies are non-transitive in that $%
R>S>P>R.$ A payoff of $+1$ has been assigned to winning, $-1$ to losing and $%
0$ for both in case of a tie.}%
\begin{tabular}{lll}
&  &  \\ 
\begin{tabular}{l}
\\ 
\end{tabular}
& 
\begin{tabular}{lll}
&  &  \\ 
&  & 
\end{tabular}
& \multicolumn{1}{|l}{$%
\begin{tabular}{lll}
& \ \ \ \ \ \ Bob \ \  &  \\ 
\ \ \ \ R & \ \ \ \ \ \ \ \ S & \ \ \ \ \ P%
\end{tabular}%
$} \\ \cline{2-3}
\begin{tabular}{l}
\\ 
\end{tabular}
& $%
\begin{tabular}{l}
\\ 
Alice \\ 
\\ 
\end{tabular}%
\begin{tabular}{l}
R \\ 
S \\ 
P%
\end{tabular}%
$ & \multicolumn{1}{|l}{$%
\begin{tabular}{lll}
$(0,$ $0)$ & $(1,$ $-1)$ & $(-1,$ $1)$ \\ 
$(-1,$ $1)$ & $(0,$ $0)$ & $(1,$ $-1)$ \\ 
$(1,$ $-1)$ & $(-1,$ $1)$ & $\ (0,$ $0)$%
\end{tabular}%
$}%
\end{tabular}%
\end{table}

\end{document}